\documentclass[11pt]{article}
  \usepackage{amssymb}
   \usepackage{graphicx}

\def\al {\alpha}

\def\ba{\begin{eqnarray}}
\def\bam{\begin{array}}

\def\be{\begin{equation}}

\def\bi{\bibitem}

\def\bt {\beta}

\def\B {\overline}

\def\Br{\B r}

\def\de{\delta}

\def\De{\Delta}

\def\ea{\end{eqnarray}} 
\def\ee{\end{equation}}

\def\fr{\frac}

\def\ha{\frac{1}{2}~}

\def\inf{\infty}

\def\la {\lambda}
\def\lb{\label}

\def\La {\Lambda}

\def\LLL{\left[}

\def\nn{\nonumber}

\def\om{\omega}

\def\Om {\Omega}

\def\ra{\rightarrow}

\def\RRR {\right]}

\def\si{\sigma}

\def\td{\tilde}

\def\ts{\textstyle}

\def\ze{\zeta}

\def\1{{\it one}}

\def\2{{\ts{\ha}\!}}

\def\3 {\ts{\frac{1}{3}\!}}
\def\4{\ts{\fr{1}{4}\!}}
\def\9{\ts{\fr{1}{9}\!}}
\begin{document}
\title{Does Viscosity turn inflation into the CMB and $\La$?}
\author{ D.Lynden-Bell$^{1}$, S.M.Chitre$^{1,2}$ 
\\  {\it$^1$  Institute of Astronomy, The Observatories,
 Madingley Road, Cambridge CB3 0HA.}\\
\it$^2$ UM-DAE Centre for Excellence in Basic Sciences,University of Mumbai,\\
\it Mumbai 400 098, India.}
\date{}      
\maketitle                                     
{\bf Abstract} 
Consideration of the entropy production in the creation of the CMB leads to a simple model of the evolution of the universe during this period which suggests a connection between the small observed acceleration term and the early inflation of a closed universe. From this we find an unexpected relationship between the Omega's of cosmology and  calculate the total volume of the universe.
\section{Introduction}
While the uniformity of the large scale structure of the Universe and the Gaussian structure of its fluctuations argue that some form of inflation occurred (See e.g.Kazanas
(1980)) ({\bf 1}), there is as yet no agreed prediction for the amplitude of the fluctuations or for the entropy of the CMB. This entropy is by far the dominant contribution to the entropy of the Universe (Basu and LB)({\bf 2}) unless the rather different gravitational entropy of giant black holes is considered. Such black holes are thought to be formed much later than the CMB as observed by the Planck satellite.
	Here we shall follow Padmanabhan and Chitre (1987)({\bf 3}) in using a purely phenomenological	fluid mechanical approach to the generation of the CMB's entropy which must have occured close to the end of the inflationary era. At that time the major contributions to the density were the density of inflationary material $\rho_i$ and the density of radiation $\rho_r$  so $\rho=\rho_i+\rho_r.$ These are associated with pressures $p_i=-\rho_i c^2$ and $p_r=\3 ~\rho_r c^2$. In a homogeneous medium in uniform expansion shear viscosity has no effect so the source of dissipative entropy creation is bulk viscosity which is zero for pure radiation by itself. We shall consider that the CMB has all the entropy which increases as a result of the decay of the inflationary density.  Let $a(t)$ be the scale factor and let $V=\3 ~4\pi a^3$. The expression for the bulk viscous stress tensor is $T^i_{j (visc)}=\ze (\de^i_j-u^i u_j)u^k_{;k}$ where $\ze $ is our phenomenological coefficient of bulk viscosity which must vanish for pure radiation, $\rho_i=0$.
We shall also suppose that it vanishes when there is only inflationary material, $\rho_r=0$.
Without viscosity the radiation-and-Lambda universe can be integrated even with k the curvature term included ({\bf 4}). Einstein's equations with the inclusion of the viscous term are 
\be
(\dot{a}^2+kc^2)/a^2=\3 ~ 8\pi G\rho,~~~~~~~~\dot{V}^2/V^2\doteq 24\pi G \rho=\om^2 X^2,
\ee
( which is unaffected by viscosity) and  which takes the second form when $k=0$ or is inflated away, where $\om^2$ is the constant $24\pi G\rho_0$ and $X^2=\rho/\rho_0$. We can leave the value of the constant density $\rho_0$ to be decided later. The other Einstein equation for homogeneous Cosmology is
\be
T dS/dt=d/dt(\rho c^2 V)+pdV/dt=\ze \dot{V}^2/V.
\ee
Here $T$ is the temperature of the radiation and $S$ is the entropy in volume $V$.
We treat our system of inflationary density $\rho_i(t)$ and radiation $\rho_r(t)$ as
analogous to a two fluid system. Neither fluid treated in isolation has any bulk viscosity but we are going to argue that there is an effective bulk viscosity for the combined two fluid system. We argue by analogy to a system of two perfect inviscid gases, one with an adiabatic index of say five thirds and the other of four thirds. If there were no interaction at all and the both  were initially at the same temperature then after expansion the fluid with the higher index would have a lower temperature than the other. The resultant heat flow would cause an entropy increase unless the whole expansion were so slow that the temperature kept them  in equilibrium at all times. In practice they would not quite equilibrate and the entropy creation via bulk viscosity depends on  $\dot{V}^2$  so it only dissipates significantly when changes are quick enough ({\bf 5}).\\
We do not claim that the system of inflationary density and radiation is a direct analogue of the two fluid system just discussed. We merely cite the latter as an example in which entropy is created via expansion and therefore via a bulk viscosity also, despite each fluid being inviscid in isolation. We shall therefore assume that our combined system has an effective coefficient of bulk viscosity $\ze$ which must vanish when either $\rho_i$ or $\rho_r$ is zero. In the example just cited there was an exchange of internal energy between the two fluids that resulted from the expansion and led to the entropy increase. Thus we are led to the idea that the effective bulk viscosity is the vehicle through which the inflationary density decays to make the CMB. Thus we set
\be 
Vd(\rho_i c^2)/dt=-\ze\dot{V}^2/V,
\ee
 where $\ze$ must vanish when either $\rho_r$ or $\rho_i$ are zero. Since the velocity field on which the viscosity acts is only defined for the radiation field (Inflationary material having no rest frame), it is natural to define a kinematic viscosity via the density $\rho_r$. For the dependence on $\rho_i$ we take the ansatz,
 \be
 \ze=\nu_0 \rho_r\sqrt{\rho_i/\rho},
 \ee
 where $\nu_0$ is a kinematic bulk viscosity with dimensions $[L^2T^{-1}]$. The square root dependence leads to easier mathematics. In practice we shall work with a dimensionless kinematic viscosity $\nu=\frac{3}{4}\nu_0~\om/c^2$, so then $\ze\dot{V}^2/V^2=(4/3)\nu c^2\om \rho_r X Y$. Indeed with our final choice of $\rho_0$ as the density of the initial inflation our dimensionless $\nu$ will turn out to be one but for the present we retain it and leave $\rho_0$ to be decided later. We have now completed the physical input on which our model is based. Notice that we have only postulated something with the equation of state $p_i= -\rho_i c^2$. We have not postulated an scalar field with a specially chosen potential but we have assumed that inflation decays and we have postulated that it does so via the effective bulk viscosity that appears when two substances exchange energy via the expansion. Since the laws of thermodynamics are universal our only special assumption is the form assumed for the bulk viscosity. For other studies of inflation's end see e.g. ({\bf 6}). 
 \section{Mathematical solution and elucidation} 
 Writing $Y^2=\rho_i/\rho_0$ equation (3) divided by $ Vc^2\rho_0$ becomes
\be
dY^2/dt=-[\ze/(\rho_0 c^2)]\dot{V}^2/V^2
\ee
and equation(2) divided by $\rho_0c^2 V $ becomes
\be
dX^2/dt=-4/3(\rho_r/\rho_0)\dot{V}/V+[\ze/(\rho_0c^2)]\dot{V}^2/V^2
\ee
 On division and simplification
\be
\frac{XdX}{YdY}=\frac{\om X}{\nu Y} \frac{V}{\dot{V}}-1,~~~~~~~~~(\frac{1}{\nu}-Y)\frac{dY}{dX}=X,
\ee
where we used the approximate  form of equation (1) only to get the final equation.
 Now when the total density is zero, the inflationary density must be zero too,  so on integration we get
\be
(2/\nu)Y-Y^2=X^2,
\ee
so the normalised radiation density $\rho_r /\rho_0=X^2-Y^2=2Y(\nu^{-1}-Y)$. We now return to equation (5) and define a dimensionless time $\td t$ as follows
\be
\frac{dY}{[Y(\nu^{-1}-Y)]\sqrt{2\nu^{-1}Y-Y^2}}=-\3 ~4\nu\om dt=-\nu^2 d\td t.
\ee
Writing $Y=\nu^{-1}(1-\cos\phi)$ and then $\tan(\phi/2)=\tau$ the integral reduces to $\nu^2$ times
\be
\int\frac{d\phi}{(1-\cos\phi)\cos\phi}=\int\frac{(1+\tau^2)d\tau}{\tau^2(1-\tau^2)}=-\frac{1}{\tau}+\ln(\frac{1+\tau}{1-\tau}),~~~~Y=\frac{2\tau^2}{\nu(1+\tau^2)}.
\ee
so 
\be
\td t(\tau)=\frac{1}{\tau}+\ln(\frac{1-\tau}{1+\tau})=\frac{16c^2 t}{9\nu_0}= \frac{4\om t}{3\nu},~~X=\frac{2(\tau/\nu)}{1+\tau^2}, ~~\frac{\rho_r}{\rho_0}=X^2-Y^2=\frac{4\tau^2(1-\tau^2)}{\nu^2(1+\tau^2)^2}.
\ee
The graph of $\td t(\tau)$ is shown in figure 1 .
 \begin{figure}[htbp]
\begin{center}
\includegraphics[width=15cm]{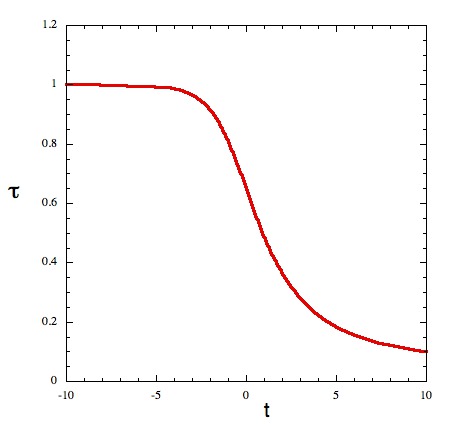}
\caption{ $\tau$ as a function of time in suitable units. For a long time from $-\inf~~~\tau$ remains very close to one as the universe inflates. Then viscous dissipation turns inflationary energy into the CMB. Thereafter $\tau \propto 1/\td t$ which is the era of the radiation universe.}
\label{fig1}
\end{center}
\end{figure}
As $\tau$ tends to unity from below, $t$ tends to negative infinity, $X$ and $Y$ become $1/\nu$,
the radiation density vanishes thus at negative infinite time we start in a pure inflationary state
with a density $\rho_I=\rho_0/\nu^2$.  We now choose $\rho_0$ to be the density of this initial inflationary state. This implies that our dimensionless $\nu=1$ and $\om^2=24\pi G\rho_I ~~~\nu_0=\3 ~4c^2/\om$.The $\nu$s in the denominators of $X,Y,X^2-Y^2, etc.$ all disappear.\\ From nearly infinite negative time to  a finite but negative time, $\tau$ decreases very gradually from one and when $(1-\tau)/(1+\tau)=1/e^2$ i.e. $\tau=(e^2-1)/(e^2+1)=0.76$ the logarithmic term becomes $-2$ whereas it was negatively infinite. $\td t =-(e^2-3)/(e^2-1)=-~0.689$. Later when  $(1-\tau)/(1+\tau)=1/e,~~~
\tau =(e-1)/(e+1)=0.462$ and $\td t=+2/(e-1)=1.164$. Soon thereafter the logarithmic term has little influence and the standard solution for the radiation filled universe takes over with $\td t=1/\tau$ and $\3 ~8\pi G\rho=1/(2t)^2~~~~~~\rho_r=\rho.$
A nice feature of this model is that we can integrate to find the scale factor. From equations (1) and (9)
we deduce that
\be
\fr{\dot{a}}{a\dot{Y}}=\fr{da}{a ~d Y}=\fr{\om X}{3\dot{Y}}=- \frac{1}{2\nu(X^2-Y^2)},
\ee
which thanks to equation (8) integrates to give
\be
(a/a_h)^4=[(\nu Y)^{-1}-1]=(1-\tau^2)/(2\tau^2)
\ee
where $a_h$ is the value of $a$ when $\nu Y=1/2$ and the radiation density reaches its maximum of $\rho_I/2$.  At large times $1/\tau= \td t,~~a\propto \td t^{1/2}$ and we have the radiation universe. From now on we set $\rho_0=\rho_I$.
From equations (13) and (11)
\be
\rho_r a^4=2\rho_I a_h^4 \LLL\frac{1-\tau^2}{1+\tau^2}\RRR^2\ra2\rho_Ia_h^4
\ee
which becomes constant as $\tau\ra 0$.This gives the energy in the black body radiation
at late times such as the present. Using $\si$ for Stefan's constant, standard statistical mechanics of black body radiation gives an entropy density $s=(4/3)(4\si/c)T^3$, corresponding to an energy density $ \rho_r c^2=(4\si/c)T^4$, thus $s=(4/3)(4\si/c)^{1/4}(\rho_r c^2)^{3/4}$. The entropy per unit comoving volume merely accumulates as the inflationary matter decays. The expansion does not affect it. The total entropy generated in the closed universe at the end of the inflationary era is the entropy density times the volume of the closed universe as measured today. This is evaluated at $\tau=0$.
\be
S=2\pi^2a^3s=(8/3)(4\si/c)^{1/4}\pi^2[a^3(\rho_r c^2)^{3/4}]_0;~~~~~~k=+1
\ee
We know all the terms except $a$. 
\begin{figure}[htbp]
\begin{center}
\includegraphics[width=15cm]{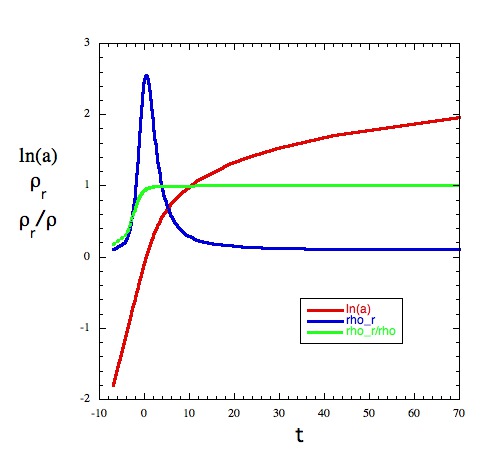}
\caption{ The radiation density  both linearly and as a fraction of the total density is plotted together with $\ln(a)$ as functions of time $\td t$. As initial inflation ends there is a strong peak in the radiation density whose maximum reaches a half of the initial inflationary density $\rho_I$.}
\label{fig2}
\end{center}
\end{figure}
\section{Late re-inflation in a closed universe}
In the considerations above we neglected the curvature herm $k/a^2$ in equation (1). It is indeed small near the end of the inflationary era and in the radiation era but both the inflationary density and the radiation density eventually become small. Could it be that this small term eventually outlasts the others? We shall evaluate this small term using perturbation theory. The second equation (1)
becomes
\be
\dot{V}^2/V^2=\om^2X^2[1-9kc^2/(\om^2X^2a^2)] 
\ee
Then the first of equation (7) becomes
\be
\frac{X dX}{Y dY}=\frac{1}{Y}[1+9kc^2/(2\om^2 X^2 a^2)]-1
\ee
We now multiply by $2Y$ and integrate under the initial condition that $X=Y=1=\tau$ to find
\be
X^2=2Y-\frac{9kc^2}{\om^2\ a_h^2}\int_Y^1\frac{ a_h^2 dY}{a^2 X^2}-Y^2
\ee
We evaluate the integral using the unperturbed functions $a/a_h$ given in (13) and  X given in (8) remembering that $\nu$ is now one.
The integral becomes $I=\int_Y^1(Y-Y^2)^{-1/2}(2-Y)^{-1} dY$.  We put $1/2-Y=(1/2)\cos\td\phi$ and then $\tan(\td\phi/2)=\td\tau$ and find $I=2^{-1/2}[\pi/2-\tan^{-1} (\td\tau/\sqrt 2)]$.  Note that $\td\tau=\sqrt2 ~\tau/\sqrt{1-\tau^2}$. Thus
\be
X^2=2Y-Y^2-2\de(\tau),~~~~~~~\de(\tau)=\frac{9kc^2}{\sqrt{2}~\om^2a_h^2}\LLL\frac{\pi}{2}-\tan^{-1}\frac{\tau}{\sqrt{1-\tau^2}}\RRR
\ee
Inflation ensures that $3c/(\om a_h)<< 1$; much is beyond the current horizon at the 
end of inflation so this $\de$ is indeed small.
We notice that $\de(\tau)$ does not vanish at late times when $\tau\ra 0$. 
Indeed the big bracket becomes $\pi/2$.
The destruction of inflationary density ceases as the radiation declines toward zero so we look to times when $\rho_r/\rho_I=X^2-Y^2=0$. From (19) this occurs when $\2-Y=\2 ~\De,~~~~\De^2=1-4\de$ or approximately when $Y=\de$. Thus in such a model the inflationary density never quite disappears but survives the radiation era to emerge as a Lambda-like term as the radiation dies  out.
 $\de$ has a very minor effect on the transition from the inflationary era to the radiation era but, as the radiation density finally decreases, it becomes constant and  then represents the survival of a remnant of the inflationary density. Although $\de$ is a function of $\tau$ it only becomes important when it is constant. For this reason we shall treat it as though it were constant in our mathematics but we shall remember that it is  near zero during the inflationary era but almost  constant during the radiation era.The radiation density  $X^2-Y^2=0$ when $-\de+Y-Y^2=0,~~~Y=2\de/(1\pm \sqrt{1-4\de})\doteq 2\de/[1\pm (1-2\de)]$. The upper sign gives a root close to $\de$ in place of zero while the lower sign gives a root  $1$. Any solution for $Y$ that starts between these roots is limited to lie between them at all times so the lower root gives a lower bound to the inflationary density.
If we have a very small $\de$ we may identify the small root with the current Lambda term by putting $\de=(\sqrt{\rho_\La/\rho_I}-\rho_\La/\rho_I)$. Where the final term is negligible. Using (19) and then (14) and (15) this gives us an important relationship between the current $\La$-term, the initial inflation in a closed $k=1$ universe and the Total entropy in its CMB currently $(\tau\ra\de)$.
\ba
\sqrt{\La c^2}=\sqrt{8\pi G \rho_\La}=\frac{(3\pi/2)kc^2}{\sqrt{16\pi G \rho_I} a_h^2}=\frac{(3\pi/2)kc^2}{(\sqrt{8\pi G \rho_r}
a^2)_0}=C_1/S^{2/3}; \\
\sqrt{\Om_\La \Om_r}= - (\pi/2)~\Om_k
\ea
where $C_1$ is a known constant and (21) follows from the definitions of the $\Om s$ with $\Om_k=-k c^2/(a^2H_0^2)$. Notice that a negative $k$ is inconsistent with (21)
so on this theory renewed inflation is evidence for a closed universe. For $\Om_\La=0.7$ and $\Om_r=8.8 ~10^{-5}$ equation (21) gives $\Om_k= -0.0050$. This is fully consistent with the Planck data on its own ({\bf 7}) which gives $-0.005\pm.017$ and is also  consistent with that data supplemented with data from Baryon Acoustic Oscillations which gives $ 0.00\pm.005 $.
From equation (20) we may determine the total entropy from the measured $\La$ term
and dividing the total entropy by the measured entropy density we find the volume of the whole universe. We can actually do this more directly by solving equation (20) for $a^2$ and using the expression $V_T=2\pi^2 a^3 = 55777 (c/H_0)^3$ for the total volume of the closed universe and  $a=14.1c/H_0$.\\
For small delta the expression for $\td t(\tau)$ is only slightly modified to
\be
\td t(\tau)=\frac{1}{q^3\tau}+\ln(\frac{1-\tau}{1+\tau})+O(\de^2)=\frac{16c^2 t}{9\nu_0}= \frac{4\om t}{3},~q^2=1-2\de,  
\ee
 When we do not neglect $\de^2$ the expression for $\td t$ is more complicated but 
 putting $\de=0$ for the terms that dominate during inflation and leaving it for the terms that matter later we find
\ba
\td t(\tau)=\ln\frac{1-\tau}{1+\tau}+\ln\LLL\frac{\tau+\bt}{\tau-\bt}\RRR^{\mu}~~~\mu=\frac{1-\De+2q}{\De\sqrt{( q+q^2+\de)(q^2-\de-q\De)}} \doteq 2/\de,\\\bt^2=\frac{\de^2}{2(q+1-\de)}\doteq\LLL\frac{\de}{2}\RRR^2,~~~~\De^2=1-4\de,~~~~q^2=1-2\de,\nn
\ea
$\tau$ now lies in the range $\beta\le \tau\le1$ and at both ends ~~$X=Y$. Notice that $\bt=O(\de^2)$. When $\tau>>\bt$ and $\de<<1$ (22) agrees with (23). The expressions for $X$ and $Y$ are {remembering that $q=1$ when $\tau=1$),
\be
Y=[1-q\frac{1-\tau^2}{1+\tau^2}]=\frac{[(2-\de)\tau^2+\de]}{(1+\tau^2)},~~~~~X=2q\tau/(1+\tau^2).
\ee
The initial inflation occurs as $\tau$ approaches one from below.  In the end
as $\tau\ra \beta,~~\td t\ra\inf\\~ Y\ra 1-q(1-\bt^2)/(1+\bt^2)\dot=\de$ so a small part of inflationary material now survives. Nevertheless there is still a long period in which radiation dominates 
\be
\rho_r=\rho_I(Y-Y^2-\de);~~~~\rho_r/\rho=(X^2-Y^2)/X^2=\frac{ Y-Y^2-\de}{Y-\2 Y^2-\de}.
\ee
Whenever $ Y$ is small but much greater than the very small $\de$, we see that the radiation is the dominant contribution to the density and from (22)  this occurs when $\tau$ is small and $\td t$ large but not so large that  the residual $\rho_\La$ takes over. We can again integrate equation (12) with the new formula for $X^2-Y^2$. Without assuming $\de$ small this gives
\be
(a/a_1)^4=\LLL\frac{ \De}{Y- (1-\De)/2}-1\RRR^{1/\De};~~~~~~~~~~~~~~\De^2=1-4\de.
\ee
 The final exponential growth of $a$ can not be recovered from this general formula and equation (22), because the exponent involves $t\de$ and we have neglected $\de^2$ in deriving (22), however once we realise that the density becomes $\rho_\La$  we can then  integrate $\dot{a}/a=\sqrt{\3 ~8\pi G\rho_\La}$ to give the  correct late time behaviour as may be checked more laboriously using  (23). It is no longer true that the total density and the inflationary density vanish together as in equation (8). This is because with the renewed inflation neither ever vanish.\\
 The present results are crucially dependent on our boundary condition that setting $\nu=1$ the initial $\rho/\rho_I=\rho_i/\rho_I=1$ that is $X^2=Y^2=1$ initially. This could be upset e.g. by an arbitrary change of $\nu$ from that found in section 2.
 To make this toy model more realistic, dark matter and baryons must be added as a small extra decay product. When relativistic those will have a gamma of 4/3 so they will only change the model later as they become non-relativistic but the final re-inflation will be unaffected.
 \section{Conclusions}
 The exchange of energy between fluids in expansion is accomplished by an effective bulk viscosity which generates the entropy increase. The incorporation of these ideas into Cosmology leads to a simple and calculable model of the inflationary and the radiation eras which, by keeping the small curvature term for a closed universe, is readily extended to give an eventual re-inflation. The value of Lambda is related to properties of the universe in the much earlier inflationary era by $\rho_\La=[(\pi/2)( 3c^2)/(8\pi G)]^2/(\rho_I a_h^4)]$ where $a_h$ is the scale factor when half the inflationary density remains. This is re-expressed, cf equation (20), to give the value of the current Lambda term in terms of the current total entropy in the universe's CMB. 
 This has allowed us to calculate the total volume of the closed universe to be far larger than the currently observable volume within the horizon. It has also given the
 relationship $\sqrt{\Om_\La \Om_r}= - (\pi/2) \Om_k$ which can be checked observationally. In the model given the density and temperature of the CMB tend to zero as  time $t\ra-\inf$. However Padmanabhan ({\bf 8}) has claimed that the temperature should never be less than the horizon temperature so our model should perhaps start when that equality holds.\\ 
 \section{Acknowledgements}
 We benefitted greatly from the enthusiasm of Colin Norman for our pursuit of this work
 and from George Efstathiou's mastery of the Planck data. \\
References\\
({\bf 1}) Kazanas, D. 1980, ApJ. ${\bf 241}$, L59;   Guth, A. H. 1981, Phys Rev D ${\bf 23}$, 347\\
({\bf 2}) Basu B., Lynden-Bell D. 1990 Q.Jl.R.Astr.Soc. {\bf 31}, 369.\\
({\bf 3}) Padmanabhan T.,   Chitre, S.M. 1987, Phys Lett. A {\bf 120}, 433.\\
({\bf 4}) Lynden-Bell D.,  Bicak J. 2016 C Q Grav.{\bf 33}, 5001.\\
({\bf 5}) Weinberg, S. 1981, Gravitation and Cosmology, Wiley, New York p 568.\\
({\bf 6}) Clifton T., Barrow J. D. 2017  Gravity and the quantum Eds. Bagla J.S.\& Engineer S. Springer p61\\
({\bf 7})Efstathiou G., Planck Collaboration, 2016, A \& A{\bf 594},13.\\
({\bf 8}) Padmanabhan, T. 2005, Phys. Rep. {\bf 406}, 49.\\
\end{document}